\documentclass[times,11pt]{article}
\newcommand{\etal}{{\it et al.\ }}

\newcommand{\vs}{{\it vs.\ }}

\newcommand{\beq}{\begin{equation}
  \renewcommand{\int}{\intop\limits}
  \renewcommand{\oint}{\ointop\limits}}
\newcommand{\eeq}{\end{equation}}
\newcommand{\beqarr}{\par\begin{minipage}{11cm} \begin{eqnarray*}}
\newcommand{\eeqarr}{\end{eqnarray*} \end{minipage} \hfill 
   \stepcounter{equation}{\rm (\theequation)}\vspace{3mm}\linebreak}
\newcommand{\bdm}{\begin{displaymath}
  \renewcommand{\int}{\intop\limits}
  \renewcommand{\oint}{\ointop\limits}}
\newcommand{\edm}{\end{displaymath}}

\newcommand{\up}[1]{\ifmmode^{\rm #1}\else$^{\rm #1}$\fi}

\newcommand{\arcd}{\ifmmode^{\circ}\else$^{\circ}$\fi}
\newcommand{\arcm}{\ifmmode{'}\else$'$\fi}
\newcommand{\arcs}{\ifmmode{''}\else$''$\fi}

\newcounter{pagefrom}
\newcounter{pageto}
\newcounter{volume}
\newcounter{year}

\newenvironment{Titlepage}{
\vspace*{2cm}
  \begin{center}
}{
  \end{center}\par\vspace{3mm}
}

\newcommand{\Title}[1]{{\large\bf\boldmath #1 \\[3mm] {\footnotesize by} 
\\[3mm]}}

\newcommand{\Author}[2]{{\large\spaceskip 2pt plus 1pt minus 1pt #1}\\[3mm]
   {\small #2}\\[6mm]}

\newcommand{\Received}[1]{}

\newcommand{\Abstract}[2]{{\footnotesize\begin{center}ABSTRACT\end{center}
\vspace{1mm}\par#1\par
\noindent
{\bf Key words:~~}{\it #2}}}

\newcommand{\FigCap}[1]{\footnotesize\par\noindent Fig.\  %
  \refstepcounter{figure}\thefigure. #1\par}

\newcommand{\TabCap}[2]{\begin{center}\parbox[t]{#1}{\begin{center}
  \small {\spaceskip 2pt plus 1pt minus 1pt T a b l e}
  \refstepcounter{table}\thetable \\[2mm]
  \footnotesize #2 \end{center}}\end{center}}

\newcommand{\TableFont}{\footnotesize}

\newcommand{\MakeTable}[4]{\begin{table}[htb]\TabCap{#2}{#3}
  \begin{center} \TableFont \begin{tabular}{#1} #4 
  \end{tabular}\end{center}\end{table}}

\newcommand{\MakeTableSep}[4]{\begin{table}[p]\TabCap{#2}{#3}
  \begin{center} \TableFont \begin{tabular}{#1} #4 
  \end{tabular}\end{center}\end{table}}

{
\footnotesize \frenchspacing

\renewcommand{\and}{{\rm and }}
\section{{\rm REFERENCES}}
\sloppy \hyphenpenalty10000
\begin{list}{}{\leftmargin1cm\listparindent-1cm
\itemindent\listparindent\parsep0pt\itemsep0pt}}%
{\end{list}\vspace{2mm}}

\def\TYLDA{~}
\newlength{\DW}
\newcommand{\refitem}[5]{\item[]{#1} #2%
\def\REFARG{#3}\ifx\REFARG\TYLDA\else, {\it#3}\fi
\def\REFARG{#4}\ifx\REFARG\TYLDA\else, {\bf#4}\fi
\def\REFARG{#5}\ifx\REFARG\TYLDA\else, {#5}\fi.}

\newcommand{\Acknow}[1]{\par\vspace{5mm}{\bf Acknowledgements.} #1}
\usepackage{supertabular,lscape,epsfig}
\usepackage{amssymb}
\usepackage{amsmath}
\DeclareSymbolFont{ppa}{OT1}{ppl}{m}{it}
\DeclareMathSymbol{\vv}{\mathalpha}{ppa}{'166}
\begin{document}

\newcommand{\TabCapp}[2]{\begin{center}\parbox[t]{#1}{\centerline{
  \small {\spaceskip 2pt plus 1pt minus 1pt T a b l e}
  \refstepcounter{table}\thetable}
  \vskip2mm
  \centerline{\footnotesize #2}}
  \vskip3mm
\end{center}}

\newcommand{\TTabCap}[3]{\begin{center}\parbox[t]{#1}{\centerline{
  \small {\spaceskip 2pt plus 1pt minus 1pt T a b l e}
  \refstepcounter{table}\thetable}
  \vskip2mm
  \centerline{\footnotesize #2}
  \centerline{\footnotesize #3}}
  \vskip1mm
\end{center}}

\newcommand{\MakeTableSepp}[4]{\begin{table}[p]\TabCapp{#2}{#3}
  \begin{center} \TableFont \begin{tabular}{#1} #4 
  \end{tabular}\end{center}\end{table}}

\newcommand{\MakeTableee}[4]{\begin{table}[htb]\TabCapp{#2}{#3}
  \begin{center} \TableFont \begin{tabular}{#1} #4
  \end{tabular}\vspace*{-7mm}\end{center}\end{table}}

\newcommand{\MakeTablee}[5]{\begin{table}[htb]\TTabCap{#2}{#3}{#4}
  \begin{center} \TableFont \begin{tabular}{#1} #5 
  \end{tabular}\end{center}\end{table}}

\newcommand{\FigurePs}[7]{\begin{figure}[htb]\vspace{#1}
\includegraphics{#4}
\FigCap{#2\label{#3}}
\end{figure}}

\newfont{\bb}{ptmbi8t at 12pt}
\newfont{\bbb}{cmbxti10}
\newfont{\bbbb}{cmbxti10 at 9pt}
\newcommand{\uprule}{\rule{0pt}{2.5ex}}
\newcommand{\douprule}{\rule[-2ex]{0pt}{4.5ex}}
\newcommand{\dorule}{\rule[-2ex]{0pt}{2ex}}
\def\thefootnote{\fnsymbol{footnote}}
\begin{Titlepage}
\Title{The All Sky Automated Survey. The Catalog of Variable Stars.
III.~12$^{\rm\bf h}$--18$^{\rm\bf h}$ Quarter of the Southern Hemisphere}
\Author{G.~~P~o~j~m~a~{\'n}~s~k~i}{Warsaw University Observatory,
Al~Ujazdowskie~4, 00-478~Warszawa, Poland\\
e-mail:gp@astrouw.edu.pl}
\Author{Gracjan Maciejewski}{Toru\'n Centre for Astronomy, N.
Copernicus University, ul.~Gagarina~11, 87-100~Toru\'n, Poland\\
e-mail:Gracjan.Maciejewski@astri.uni.torun.pl}
\end{Titlepage}

\Abstract{
This paper describes the third part of the photometric data from the
$9\arcd \times 9\arcd$ ASAS camera monitoring the whole southern hemisphere
in $V$-band. Preliminary list of variable stars based on observations
obtained since January 2001 is presented. Over 3,200,000 stars brighter than
$V$=15 on 18,000 frames were analyzed and  10453 were found to be variable
(1718 eclipsing, 731 regularly pulsating, 849 Mira and 7155 other stars).
Light curves have been classified using the improved automated
algorithm, which now takes into account 2MASS colors and IRAS infrared
fluxes.
Basic photometric properties are presented in the tables and some
examples of thumbnail light curves are printed for reference.
All photometric data are  available over the INTERNET at\\
{\it http://www.astrouw.edu.pl/\~{}gp/asas/asas.html} or {\it
http://archive.princeton.edu/\~{}asas}}{Catalogs
-- Stars: variables: general -- Surveys}

\section{Introduction}
The All Sky Automated Survey (ASAS, Pojma{\'n}ski 1997) follows the
ideas of Paczy{{\'n}}ski (1997) of continuous monitoring of the whole sky
with small automated instruments.
With its wide-field cameras ($9\arcd \times 9\arcd$) equipped with
f200/2.8 telephoto lenses and 2K$\times$2K CCDs (Pojma{\'n}ski 2001), ASAS
measures brightness of all stars, asteroids and comets brighter than
limiting magnitude $V\sim~14$ ($I\sim~13$) every night.

Current ASAS system is located in Las Campanas Observatory (operated by the
Carnegie Institution of Washington) and
consists of four independent instruments equipped with standard $V,R,I$
filters and installed together in a small automated enclosure.

The $V$ filter system is working in the instantaneous mode. Its
data are acquired, analyzed and made immediately available over
the Internet.  As of January 2004 the ASAS Alert Service is being
tested. It detects in real-time new objects brighter then $V\sim
~12$ and substantial ( greater than 6 sigma ) brightness changes
of known stars. However, due to many artifacts in data only raw
alerts are immediately available. Necessary human verification
delays final alerts by  up to 48 hours.

Variability analysis of the $V$-band data started as soon as
reasonable amount of data had been collected. We have already presented
preliminary catalogs of variable stars
in the $0^h-6^h$ (Pojma{\'n}ski 2002) and $6^h-12^h$ (Pojmañski 2003) quarters
the southern hemisphere.
This paper contains the third part of the
analyzed data - variable stars located in the fields centered between
$12^h$ and $18^h$ of right ascension.

\section{Observations and Data Reduction}

Although current location of the ASAS instruments allows for
clear observation of 70\% of the sky, only stars with declination
$\delta < 0\arcd$ are included in this catalog due to still limited
number of measurements for northern fields.

The data reduction pipe-line used to process ASAS data was
described in details by Pojma{\'n}ski (1997). Its most important
feature is simultaneous photometry made through five apertures (2
to 6 pixels in diameter). Each aperture data is processed
separately,  so one can use data obtained with the smallest one
for the faint ($V > 12$) stars and with the largest one for the
bright ($V < 9$) objects. Substantial differences between large-
and small-aperture magnitudes indicate close neighbors in which
case large apertures should be avoided. More sophisticated methods
like profile fitting and image subtraction were tested but did not
perform well for highly variable and undersampled ASAS images.

Astrometry is based on the ACT (Urban \etal 1998) catalog and achieves
positional accuracy better than 0.2 pixels ( $<3$~arc sec).

The zero-point offset of photometry is based on the Tycho
(Perryman {\em et al.} 1997) data.
A few hundred Tycho stars are usually located in each
$9\arcd\times 9\arcd$ field
and we use them for precise offset calibration.
Due to  effects caused by non-perfect flat-fielding, lack of
color terms in transformation and blending of the stars, systematic errors as
large as several tens of magnitude could be observed for particular stars.
Differential accuracy is much better, reaching 0.01 for bright stars.

\section{Variability Search}

Primary search for variables in the third quarter of the southern
sky was done at the end of 2002, but data used for variability
classification covered already three years (since 2001) of
observation. Secondary search for variability is envisaged for the
whole southern sky after completing publication of the fourth part
of our data.

Variability analysis was similar to that performed on the ASAS-2
data (Pojma{\'n}ski 2000). Only stars with highest magnitude
dispersion (top 5 \%) and having large number of deviating
observations were subject to Analysis Of Variance test
(Schwarzenberg-Czerny 1989). Objects with AOV statistics value
larger than 10., and those that have passed long-term variability
tests: variance analysis in variable-length bins and trend
analysis (average number of consecutive observations showing the
same direction of brightness change) were visually inspected.

Out of 3,200,000 stars with sufficient number of measurements 39,000
have passed initial selection criteria and  10,500 were left after
final visual inspection.  These were subject to variability classification.

\FigurePs{13cm} {2MASS $J-H$ \vs $H-K$ colors for variable stars
from the GCVS. Only 5000 dots were plotted for clarity, but gray
contours represent their true density. The largest contour
encircles 99\% of variables. Black dots and contours surrounding
50\% of objects mark location of several classes from the GCVS
catalog. }{fig1}{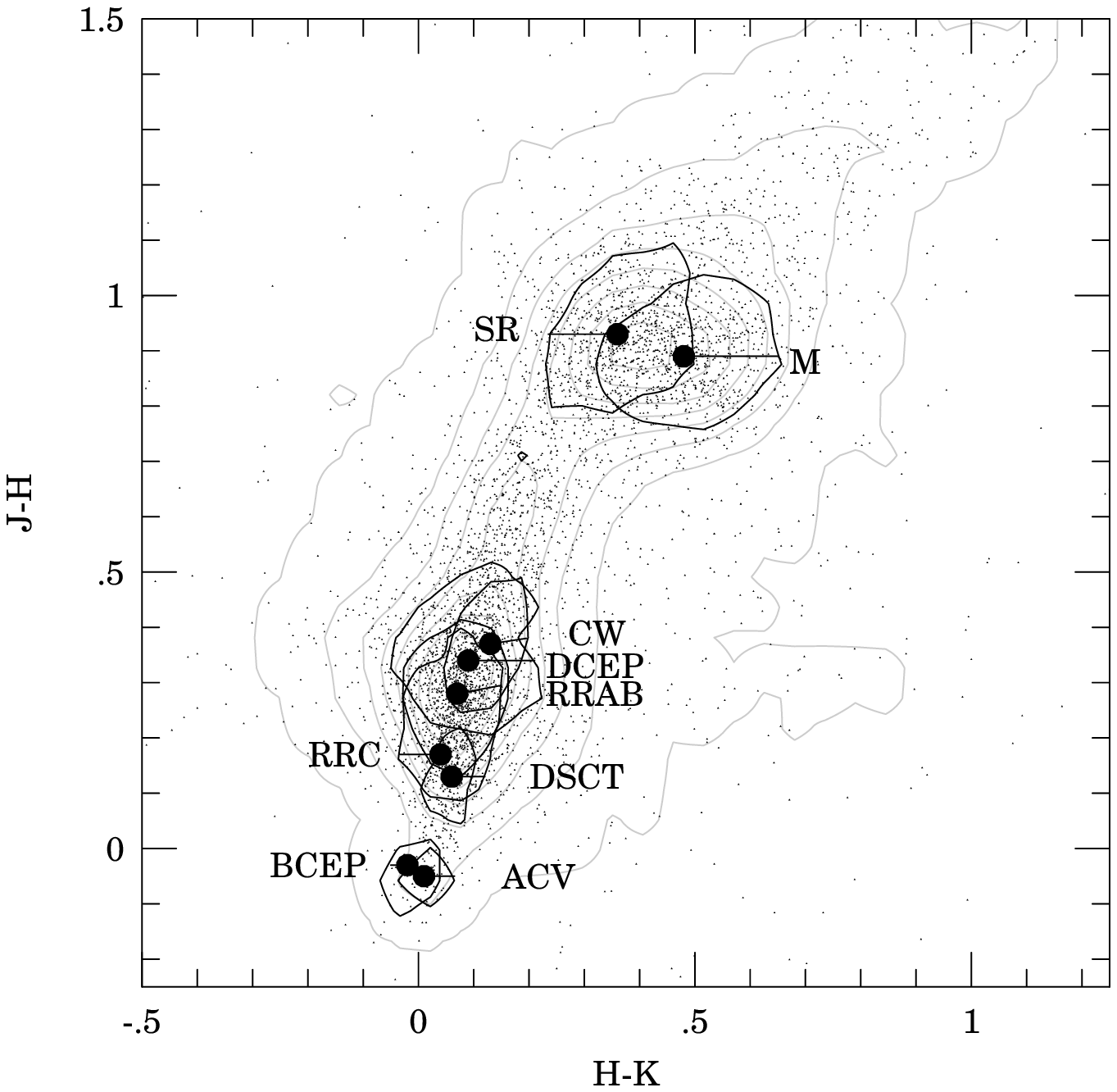}{95}{-25}{-15}

\FigurePs{13cm}
{Logarithm of period \vs 2MASS $J-H$ color for variable stars from the GCVS.
All symbols as in Fig. 1.
}{fig2}{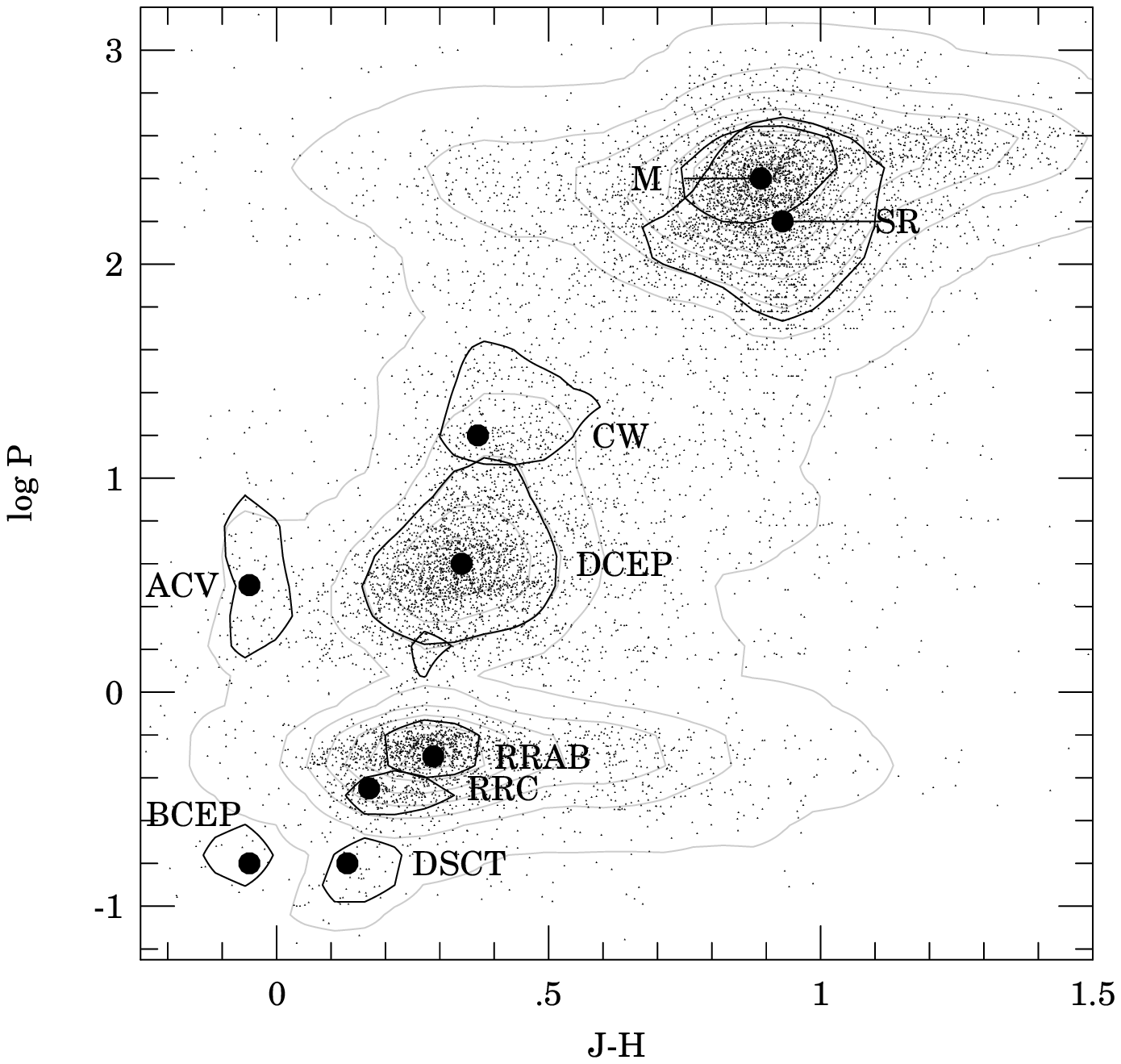}{95}{-25}{-15}

\FigurePs{13cm} {2MASS $J-H$ \vs $H-K$ colors for ASAS variables
in the third quarter of the Southern Hemisphere. All contours are
the same same as in Fig. 1. Filled and empty circles are
for Mira and $SR$ stars, filled and empty triangles for $DCEP$ and
$CW$ stars, filled and empty squares for $RRAB$ and $RRC$, crosses
for $DSCT$, pluses for $BCEP$ stars for $ACV$, tiny dots for
$MISC$, respectively. }{fig3}{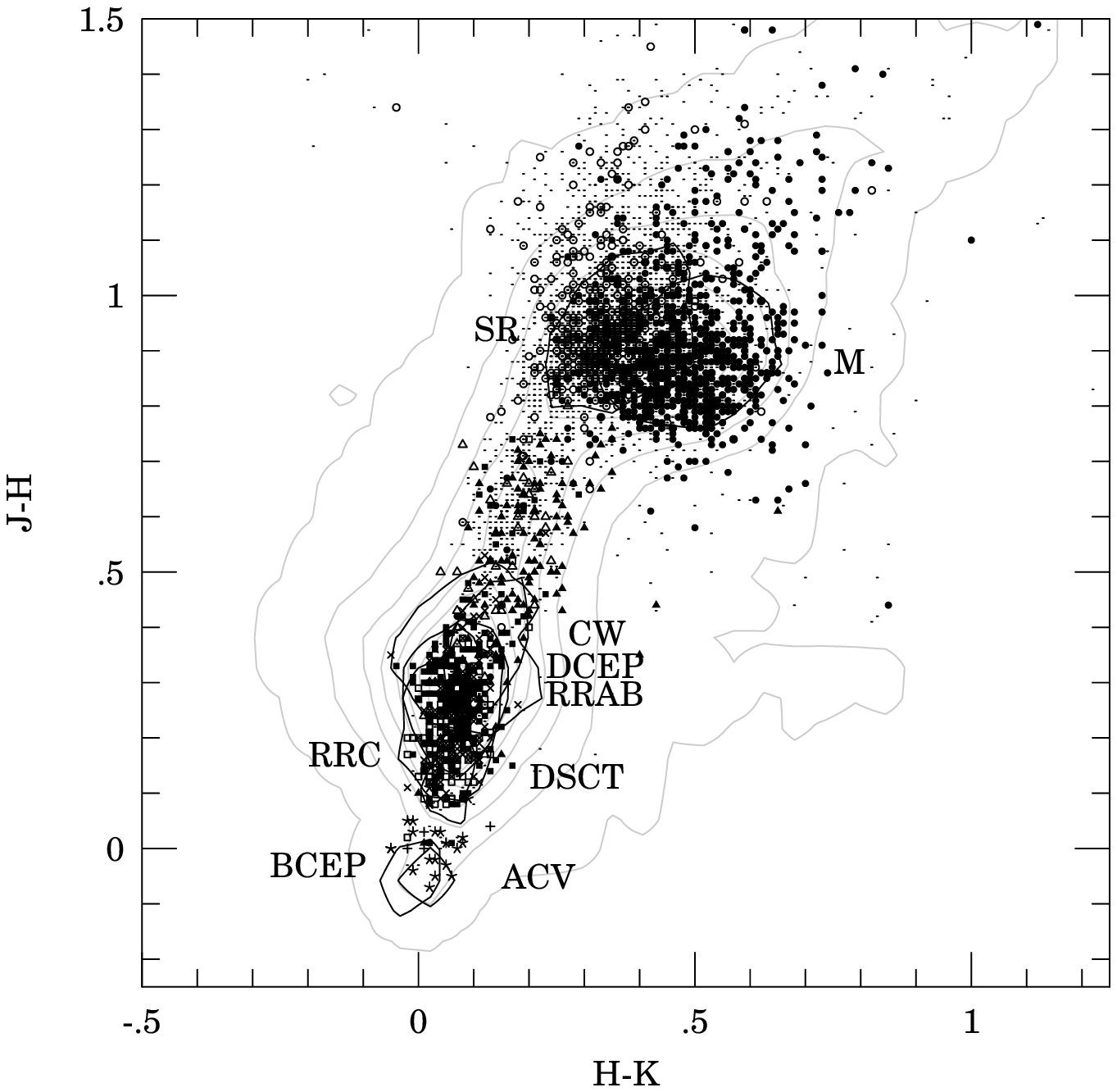}{95}{-25}{-15}

\FigurePs{13cm} {Logarithm of period \vs 2MASS $J-H$ color for
variables in the third quarter of the Southern Hemisphere. All
contours are the same same as in Fig. 2 and symbols as in
Fig. 3. }{fig4}{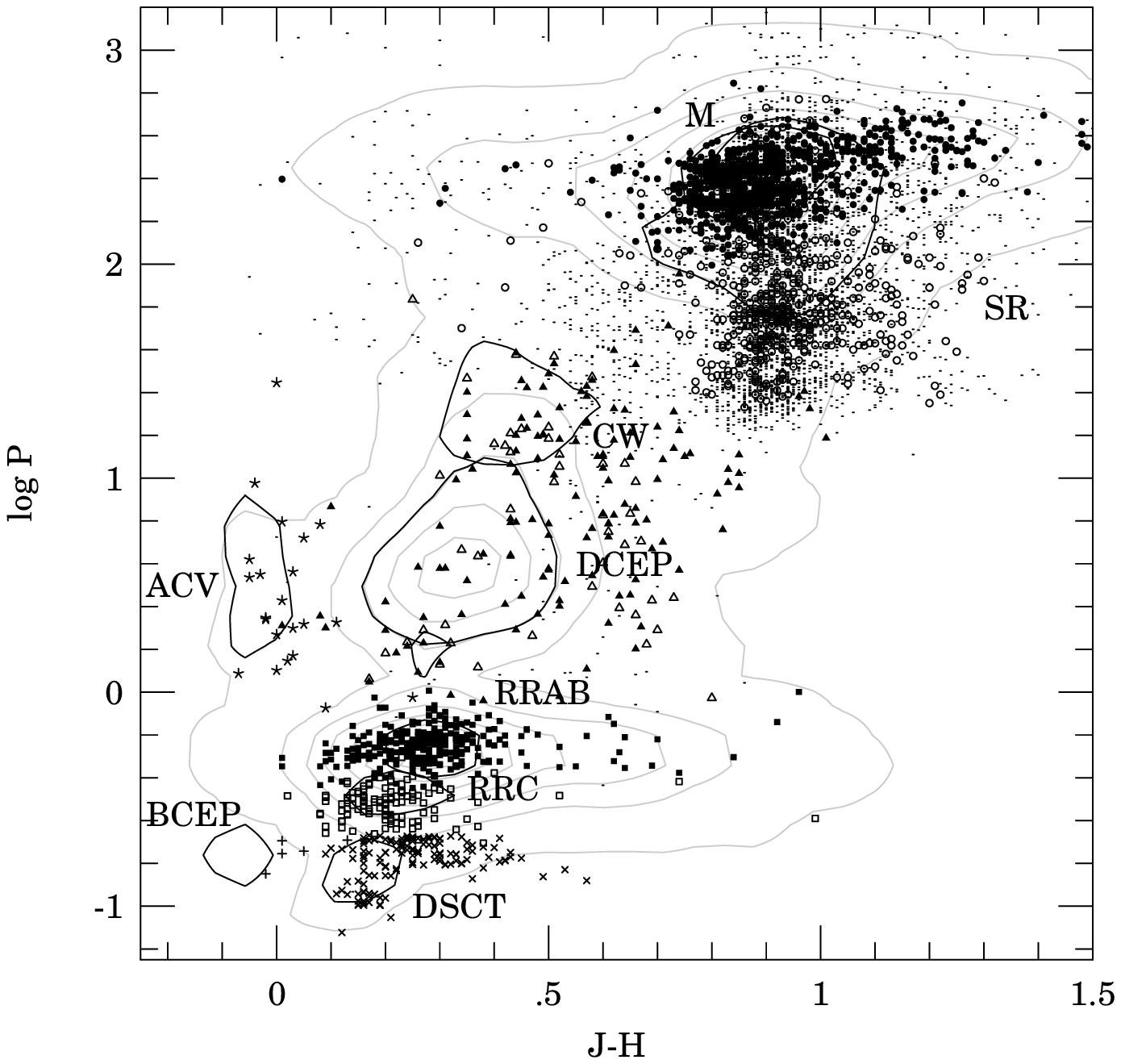}{95}{-25}{-15}

\section{Variability Classification}

Preliminary classification follows strategy described in the first and second
part of the catalog (Pojma{\'n}ski 2002, 2003).

All stars are first divided into two groups: strictly periodic and
less regular ones. This is partly accomplished by a dedicated
filter detecting difference between the actual light curve and
averaged (folded) one and partly by visual inspection.

Automated classification scheme applied in our previous analysis used
several parameters of the light curve (period, amplitude,
Fourier coefficients) to
divide stars into three eclipsing classes: $EC$
(contact or almost contact configurations),
$ED$ (detached binaries), $ESD$ (semi-detached systems)
and six pulsating categories: $DSCT$ (delta Scuti), $RRAB$, $RRC$ (RR Lyrae),
$DCEP_{FU}$, $DCEP_{FO}$ (Cepheids in fundamental and first overtone mode),
$M$ (Mira stars). All other stars ($SR$, $IRR$, $CV$, $LBV$,
Novae, many multi-periodic, etc.) have been classified as MISC.

Such approach leads to serious ambiguity if no external information is
provided. Fortunately 2MASS All-Sky Point Source Catalog is now available
and its limiting magnitude is sufficient to provide very accurate,
one-epoch $J,K,H$ measurements for most, if not all, ASAS objects.
Therefore we have added two additional parametric planes:
$J-H$ \vs ${\rm log}P$ and $H-K$ \vs $J-H$ to our classification automate.

In Fig. 1 we have plotted $H-K$ \vs $J-H$ colors for all
non-eclipsing variable stars in the GCVS (Kholopov, \etal 1985)
catalog and in Fig. 2 stars having known periods were
plotted ind $J-H$ \vs ${\rm log}P$ diagram. In both diagrams gray
contours encircle 99\%, 80\%, 70\%, ... etc. of all known GCVS
variables. Black contours labeled with variability class names
encircle 50\% of stars belonging to the class, and black dots mark
maxima of distributions.

Large scatter of color indexes is partly due to inaccurate  coordinates
in the GCVS catalog, often leading to wrong cross-identification with 2MASS
objects, and partly due to high reddening close to the galactic plane.

Figs. 1 and 2 clearly show, that 2MASS colors combined with
period sharply isolate  ACV and BCEP classes. In many cases they help to
remove EC/RRC/DSCT and SR/DCEP degeneracy (excluding pulsators
with very red colors).

Since neither light curve parameters nor infrared colors easily
separate Cepheids (DECP)  and V Virginis (CW) stars we have
decided to use galactic coordinates to make preliminary
distinction between them. This is rather statistical than physical
approach, and it should be replaced in future by more detailed
analysis of physical parameters.

In Figs. 3 and 4, similar to Figs. 1
and 2, we have plotted ASAS variables, overplotted on the
contours of GCVS distribution.

\section{The Catalog}

Current list of candidate variable stars in the third quarter of the
Southern Hemisphere contains 10453 stars.

For each star the following data are provided: ASAS identification $ID$
(coded from the star's $\alpha_{2000}$ and $\delta_{2000}$ in the
form: $hhmmss-ddmm.m$), period $P$ in days (or characteristic time
scale of variation for irregular objects), $T_0$ -  epoch of minimum
(for eclipsing) or maximum (for pulsating) brightness, $V_{max}$ -
brightness at maximum, $\Delta V$ - amplitude of variation, $Type$ - one
of the predefined classes: $DSCT$, $RRC$, $RRAB$, $DCEP_{FU}$, $DCEP_{FO}$,
$CW$, $ACV$, $BCEP$, $M$ and $MISC$.
We have also added $J$, $J-H$ and $H-K$ taken from 2MASS catalog.

Stars classified as MISC contain mostly semi-regular and irregular variables
detected by our algorithm but also many objects excluded from other classes
after visual inspection. Original classification was often appended after
MISC keyword.  1963  objects other then MISC have multiple classification.
For 485 cases this is exclusively due to $EC/ESD$ or $ED/ESD$ confusion,
but for 281 other this is a more
serious $EC/RRC$ or $EC/DSCT$ double classification. Quite often visual
inspection helped to remove such degeneracy.

\tabcolsep 5pt
\MakeTable{|l|r|l|r|}{8cm}{\label{tabvar}
Number of various types of variable stars detected in the third quarter of the southern sky by ASAS-3 $V$ camera.}{
\hline
\multicolumn{1}{|c|}{Type} & \multicolumn{1}{c|}{Count} & \multicolumn{1}{c|}{Type} & \multicolumn{1}{c|}{Count}\\
\hline
$DCEP_{FU}$ & 108   & $DSCT$&  116 \\
$DCEP_{FO}$ &  27   & $EC$  & 923 \\
$CW$        &  48   & $ED$  & 278 \\
$ACV$       &  22   & $ESD$ & 517 \\
$BCEP$      &   8   & $M$   &  849 \\
$RRAB$      & 283   & $MISC$& 7155 \\
$RRC$       & 118   &       &  \\
\hline
}

Search for GCVS variables revealed about 2350 possible
matches within 3 arc minute radius.

Table 1 summarizes our classification effort and
Table 2 contains a compact version of the
catalog. Only four  columns are
listed  for each star: identification $ID$, $P$, $V$, and $\Delta V$.
Column $ID$ also contains some
flags - ":" if classification was uncertain, "?" if multiple classes
were assigned (objects were grouped in the table according to the highest rank
assignment), "v" if SIMBAD lists a star to be variable.

Appendix shows exemplary light curves. Only $ID$ is given for each.
For periodic variables
phase in the range ($-0.1$ - $2.1$) is plotted along the $x$-axis, while for
Mira's and miscellaneous stars -  HJD in the range (2451800-2453400).
Larger ticks on $y$-axis always mark
1 magnitude intervals and vertical span is never smaller than 1 mag.

The full catalog of variables observed by the ASAS system,
containing more classification details, as well as complete data for the
light curves, is available over the INTERNET:\\
\centerline{ http://www.astrouw.edu.pl/\~~gp/asas/asas.html}
 or
\centerline{http://archive.pinceton.edu/\~~asas}.

\section{Conclusions}

We have presented preliminary catalog of variable stars
located between $12^{\rm h}$ - $18^{\rm h}$ in the Southern Hemisphere.
Almost 10500 variable stars were found
among 3,200,000 stars brighter than $V\sim 15$ mag.
Our catalog is still incomplete
and total number of variable stars will increase substantially
when more data are analyzed.

We have added 2MASS colors and galactic coordinates to our
classification algorithm, what resulted in discrimination between
$DCEP$, $ACV$, $BCEP$ and $CW$ pulsators and better definition of
other classes.

\Acknow{
This project was made possible by a generous gift from Mr. William
Golden to Dr. Bohdan Paczy{\'n}ski, and funds from Princeton University.  It is a
great pleasure to thank Dr. B. Paczy{\'n}ski for his initiative, interest,
valuable discussions, and the funding of this project.

I am indebted to the OGLE collaboration (Udalski, Kubiak, Szyma{\'n}ski 1997)
for the use of facilities of the
Warsaw telescope at LCO, for their permanent support and maintenance of the
ASAS instrumentation, and to The Observatories of the
Carnegie Institution of Washington for providing
the excellent site for the observations.

This research has made use of the SIMBAD database,
operated at CDS, Strasbourg, France and
of the NASA/ IPAC Infrared Science Archive, which is operated by the Jet Propulsion Laboratory, California Institute of Technology, under contract with the National Aeronautics and Space Administration.

This work was supported by the KBN 2P03D02024 grant.
}
\vspace{1.5cm}
\begin{center}
References
\end{center}

\noindent
\begin{itemize}
\leftmargin 0pt
\itemsep -5pt
\parsep -5pt
\refitem{Kholopov, P.N., \etal}{1985}{~}{~}{General Catalog of Variable
Stars, The Fourth Edition, Nauka, Moscow}
\refitem{Joint IRAS Science Working Group}{1988}{~}{~}{Infrared Astronomical Satellite Catalogs. The Point Source Catalog, NASA RP-1190}
\refitem{Paczy{\'n}ski, B.}{1997}{~}{~}{``The Future of Massive Variability
Searches'', in
{\it Proceedings of 12th IAP Colloquium}: ``Variable Stars and the
Astrophysical Returns of Microlensing Searches'', Paris (Ed. R. Ferlet),
p.~357}
\refitem{Perryman, M.A.C. \etal}{ 1997}{ Astron. Astroph}{ 323}{ L49}
\refitem{Pojma{\'n}ski, G.}{1997}{Acta Astron.}{47}{467}
\refitem{Pojma{\'n}ski, G.}{2000}{Acta Astron.}{50}{177}
\refitem{Pojma{\'n}ski, G.}{2002}{Acta Astron.}{52}{397}
\refitem{Pojma{\'n}ski, G.}{2003}{Acta Astron.}{53}{341}
\refitem{Udalski, A. Kubiak, M., and Szyma{\'n}ski, M.}{1997}{Acta Astr.}{47}{319}
\refitem{Schwarzenberg-Czerny, A.}{1996}{Astrophys. J.}{460}{L107}
\refitem{Urban,S.E., Corbin T.E., Wycoff, G.L.}{1998}{AJ}{115,1709}{2161}
\end{itemize}
\tabcolsep 3pt
\input{vartab.tex}
\textheight 24cm
\begin{figure}[p]
\vglue-3mm
\centerline{\bf Appendix}
\vskip1mm
\centerline{\bf ASAS Atlas of Variable Stars. ${\bf 12^{\bf h}{-}18^{\bf h}}$
Quarter of the Southern Hemisphere}

\centerline{\small Only several light curves of each type are printed. Full Atlas
 is
available over the {\sc Internet}:}

\centerline{\it http://www.astrouw.edu.pl/\~{}gp/asas/appendix12.ps.gz}

\vskip20mm
\centerline{Stars classified as EC}
\vskip3.5cm
\centerline{Stars classified as ESD}
\vskip3.5cm
\centerline{Stars classified as ED}
\vskip3.5cm
\centerline{Stars classified as DSCT}
\vskip-11.7cm
\centerline{\includegraphics[bb=30 70 400 495, width=13cm]{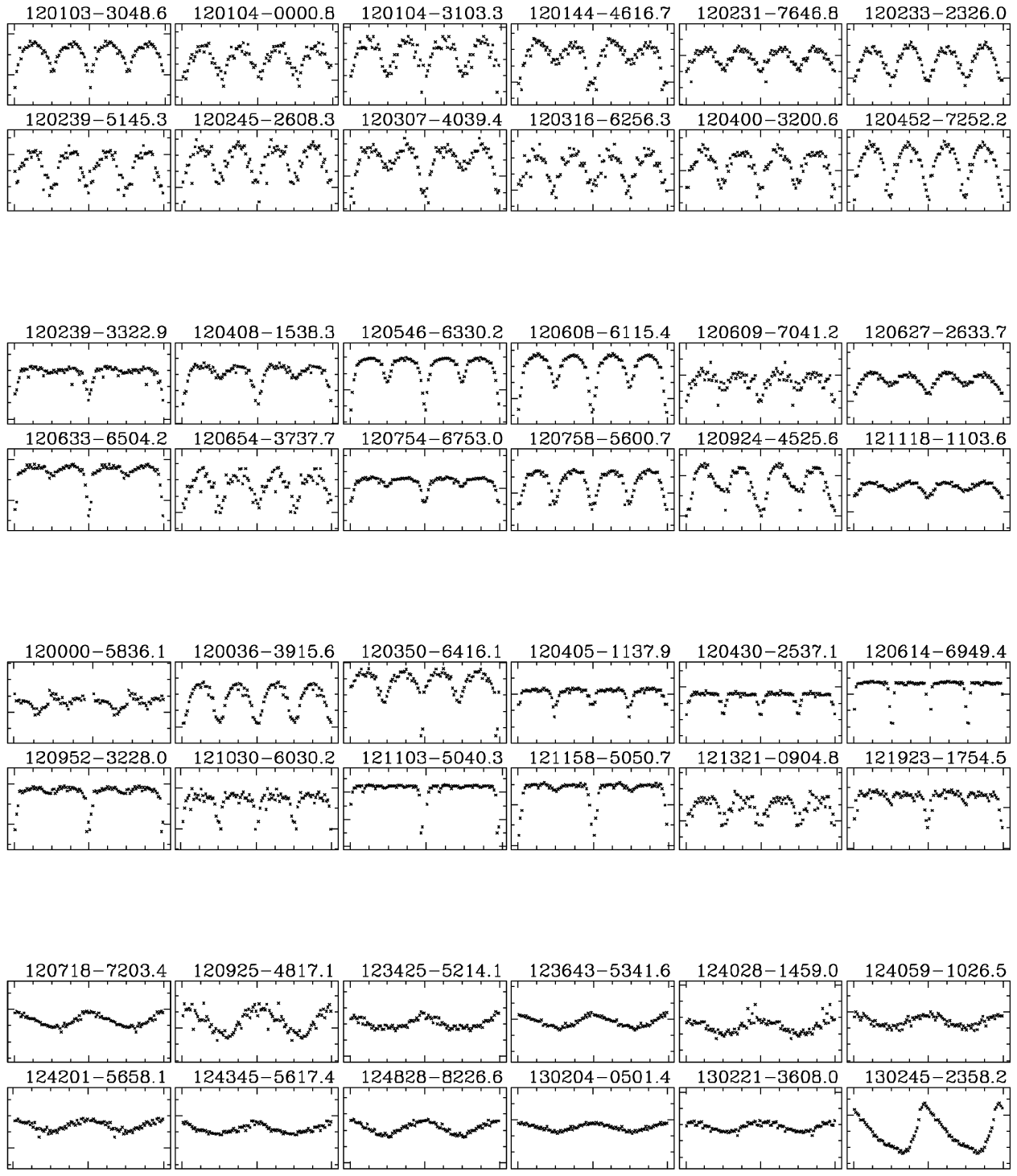}}
\end{figure}
\begin{figure}[p]
\vskip-2mm
\centerline{Stars classified as BCEP}
\vskip3.5cm
\centerline{Stars classified as ACV}
\vskip3.5cm
\centerline{Stars classified as RRC}
\vskip3.5cm
\centerline{Stars classified as RRAB}
\vskip3.5cm
\centerline{Stars classified as DCEP-FU}
\vskip-15.7cm
\centerline{\includegraphics[bb=30 30 405 610, width=13cm]{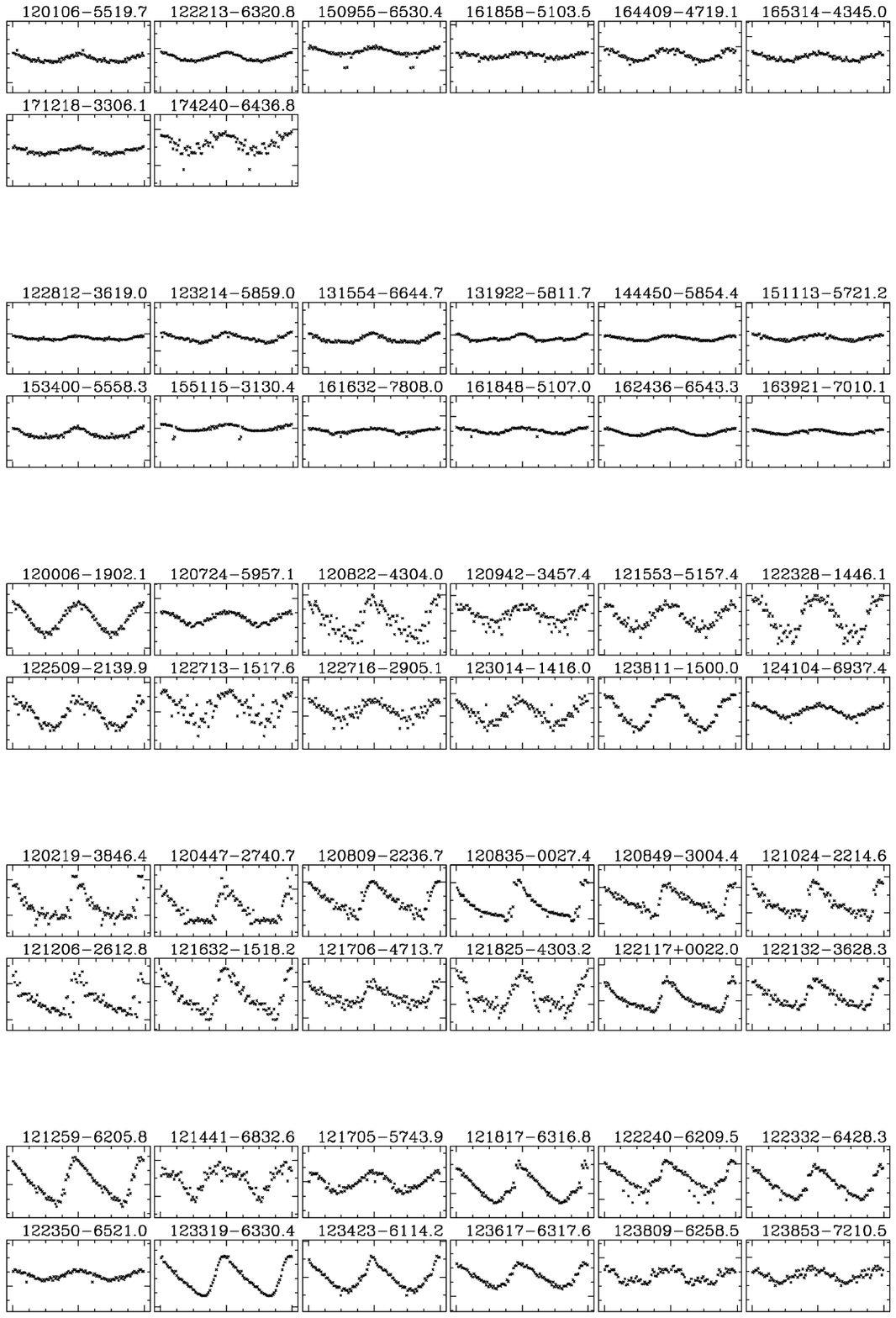}}
\end{figure}

\begin{figure}[p]
\vskip-2mm
\centerline{Stars classified as DCEP-FO}
\vskip3.5cm
\centerline{Stars classified as CW}
\vskip3.5cm
\centerline{Stars classified as M}
\vskip3.5cm
\centerline{Stars classified as MISC}
\vskip3.5cm
\vskip-15.2cm
\centerline{\includegraphics[bb=30 65 405 610, width=13cm]{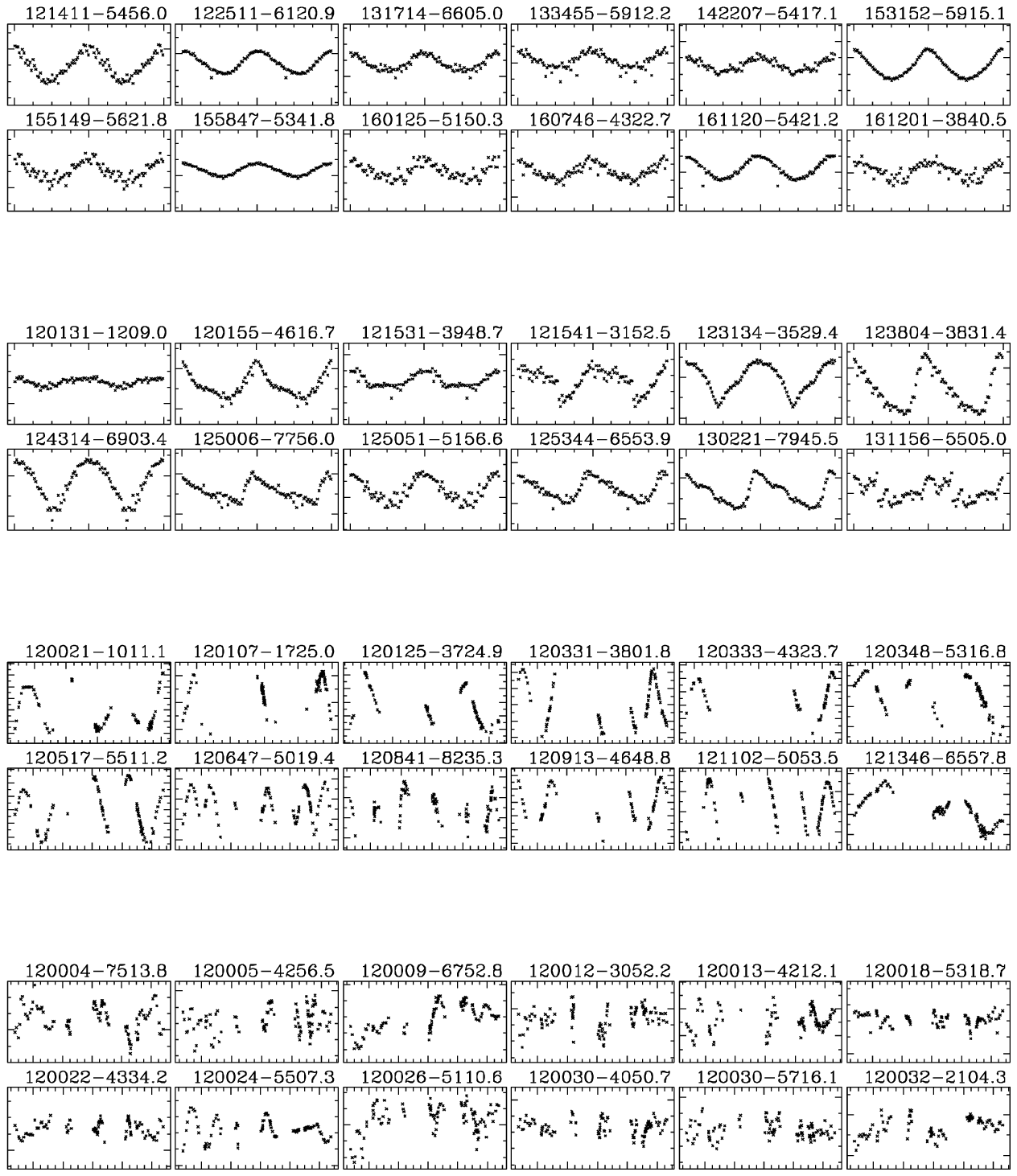}}
\end{figure}

\end{document}